\documentclass[prd,twocolumn,amsmath,amssymb,floatfix,nofootinbib]{revtex4}
\usepackage{mathrsfs,bm}
\usepackage{longtable,lscape}
\usepackage{txfonts}
\usepackage{indentfirst}
\usepackage{graphicx,color,dcolumn,booktabs}
\usepackage{multirow}
\usepackage{indentfirst}
\usepackage{multirow}
\usepackage{epsfig}
\usepackage{graphicx,color,dcolumn}
\usepackage{epstopdf}
\usepackage{amsmath}
\usepackage{multirow}
\usepackage{diagbox}
\usepackage{changes}
\usepackage{threeparttable}

\usepackage{color}
\usepackage{amssymb}
\definecolor{cover}{rgb}{0.77,0.87,0.88}
\definecolor{blueone}{rgb}{0.1,0.1,.7}
\definecolor{citec}{rgb}{0.14,0.47,0.09}
\definecolor{two}{rgb}{0.0,0.5,0.}
\definecolor{three}{rgb}{.5,.1,0.15}
\usepackage[bookmarks=true,bookmarksopen=false,plainpages=false,breaklinks=true,
   bookmarksnumbered=true,hypertexnames=false,
   filecolor=blue,urlcolor=three,menucolor=three,
   linkcolor=three,citecolor=blueone, colorlinks,
   anchorcolor=blue,runcolor=pink,frenchlinks=red
   pdfstartview=FitH,pdftitle=title,%
   pdfauthor=author]{hyperref}

\def\half{{\textstyle{1\over 2}}}

\usepackage{relsize}
\usepackage{xspace}
\def\babar{\mbox{\slshape B\kern-0.1em{\smaller A}\kern-0.1em
    B\kern-0.1em{\smaller A\kern-0.2em R}}}

\allowdisplaybreaks[4]
\usepackage{color}
\begin{document}

\title{$P^\Lambda_{\psi s}(4459)$ and $P^\Lambda_{\psi s}(4338)$ as molecular
states in $ J/\psi \Lambda$ invariant mass spectra}

\author{Jun-Tao Zhu,  Shu-Yi Kong, Jun He\footnote{Corresponding author:
junhe@njnu.edu.cn}} \affiliation{$^1$School of Physics and Technology, Nanjing
Normal University, Nanjing 210097, China}

\date{\today}
\begin{abstract}

Recently, the LHCb Collaboration has reported two strange hidden-charm
pentaquark states named $P^\Lambda_{\psi s}(4459)$ and $P^\Lambda_{\psi
s}(4338)$ in the $ J/\psi \Lambda$ invariant mass spectra of decays $\Xi_b^-
$$\to$$ J/\psi \Lambda K^-$  and $B^-$$\to$$J/\psi \Lambda \bar{p}$,
respectively.  In this work, we perform a coupled-channel study of the
interactions $\Xi_c^*\bar{D}^*$, $\Xi'_c\bar{D}^*$, $\Xi^*_c\bar{D}$,
$\Xi_c\bar{D}^*$, $\Xi'_c\bar{D}$, $\Lambda_c\bar{D}_s^{*}$, $\Xi_c\bar{D}$,
$\Lambda_c\bar{D}_s$, and $\Lambda J/\psi$ in the quasipotential Bethe-Salpeter
equation approach to estimate the $ J/\psi \Lambda$ invariant mass spectra.
With the help of effective Lagrangians, the potential kernel can be constructed
by meson exchanges to obtain the scattering amplitudes, from which the poles of
the bound states and the invariant mass spectra can be reached. The
coupled-channel calculation results in that the width of state $\Xi_c
\bar{D}^{*}(1/2^-)$ is about $18$~MeV and  that of state $\Xi_c
\bar{D}^{*}(3/2^-)$ is only about $1.6$~MeV. By comparison with experimental
data, it indicates that the structure $P^{\Lambda}_{\psi s}(4459)$ is mainly
from the contribution from  the $\Xi_c \bar{D}^{*}(1/2^-)$ state while the role
of state $\Xi_c \bar{D}^{*}(3/2^-)$ cannot be excluded.  The line shape of the
structure $P^{\Lambda}_{\psi s}(4338)$  can be reproduced roughly by a  narrow
molecular state from the $\Xi_c \bar{D}$ interaction with $J^P$=$1/2^-$, which
is extremely close to the threshold, with a large interference effect.  Besides, an
additional state $\Xi'_c \bar{D}(1/2^-)$ is suggested to be observed as a dip
structure in the $J/\psi \Lambda$ invariant mass spectrum.

\end{abstract}

\maketitle
\section{INTRODUCTION}

In the recent years, one of the most important experimental progresses of the
exotic hadron studies is the observation of $P^N_{\psi }(4450)$ and
$P^N_{\psi}(4380)$ at LHCb in 2015~\cite{Aaij:2015tga}, which were predicted by
several theoretical
groups~\cite{Wu:2010jy,Yang:2011wz,Wang:2011rga,Xiao:2013yca}. Inspired by the
experimental observation, more theoretical studies of the hidden-charm
pentaquark structures emerge~\cite{Chen:2016qju,Guo:2017jvc}.  Subsequently,
the $P^N_{\psi}(4450)$ was separated into two structures $P^N_{\psi}(4440)$ and
$P^N_{\psi }(4457)$, and a new structure $P^N_{\psi}(4312)$ was observed at LHCb
in 2019~\cite{Aaij:2019vzc}. The observations of three states with small widths
and masses close to corresponding thresholds well illustrate the validity of the
molecular state interpretation~\cite{Chen:2015loa,Chen:2015moa,
Karliner:2015ina,Roca:2015dva,He:2015cea,Burns:2015dwa}.  Furthermore, the
theoretical studies about molecular states from nonstrange hidden-charm systems
have been extent into hidden-charm pentaquark states with strangeness ~\cite{
Anisovich:2015zqa,
Wang:2015wsa,Feijoo:2015kts,Lu:2016roh,Chen:2015sxa,Chen:2016ryt,Xiao:2019gjd,Zhang:2020cdi,Wang:2019nvm}.

In 2020, the LHCb  Collaboration reported a $3\sigma$ strange hidden-charm
pentaquark structure $P^\Lambda_{\psi s}(4459)$ in the $\Xi_b^- \rightarrow
J/\psi \Lambda K^-$ decay~\cite{Aaij:2020gdg}.  This structure has a mass of
19~MeV below the $\Xi_c \bar{D}^{*} $ threshold and a width of
17~MeV~\cite{Aaij:2020gdg}, which is consistent with the properties of a
molecular state.  There are a wide variety of studies of the molecular state
interpretation of $P^\Lambda_{\psi s}(4459)$ in the literature.  In
Ref.~\cite{Wang:2021itn}, the authors considered the molecular state
interpretation for this state and concluded that it is either a $\Xi'_c \bar{D}$
state with $I(J^P)$=$0 (3/2^-)$,  or a $\Xi_c \bar{D}^{*} $ state with
$0 (3/2^-)$.  In Ref.~\cite{Chen:2020uif}, a calculation was also performed
with the QCD sum rule, and the result supports the interpretation of
$P^\Lambda_{\psi s}(4459)$ as a $\Xi_c \bar{D}^{*} $ molecular state with either
$J^P$=$1/2^-$ or $3/2^-$.  In Ref~\cite{Chen:2020kco}, it was suggested
that its two-body strong decay behavior supports an assignment of
$P^\Lambda_{\psi s}(4459)$ as a $\Xi_c \bar{D}^{*} $ state with $I(J^P)=0
(3/2^-)$. In Ref~\cite{Peng:2020hql}, the results under the heavy quark spin
symmetry (HQSS) limits  also prefer $\Xi_c \bar{D}^{*}(3/2^-) $ to  $\Xi_c
\bar{D}^{*}(1/2^-)$ as the candidate of $P^\Lambda_{\psi s}(4459)$.

Very recently, the LHCb Collaboration reported their results about the
$B^-$$\to$$J/\psi \Lambda \bar{p}$ decay, which indicates a new neutral strange
hidden-charm pentaquark state named $P^{\Lambda}_{\psi s}(4338)$. It carries a
mass of $4338.3 \pm 0.7\pm0.4$~MeV and a width of $7.0\pm 1.2
\pm1.3$~MeV~\cite{Collaboration:2022boa}.  Since its mass and narrow width are
in good line with the properties of molecular state, many new researches about
the $P^{\Lambda}_{\psi s}(4338)$ also suggest it as a molecular
state~\cite{Wang:2022neq,Wang:2022mxy,Yan:2022wuz,Ozdem:2022kei,Wang:2022tib,Nakamura:2022jpd,Giachino:2022pws,Ortega:2022uyu}.
In Ref.~\cite{Nakamura:2022jpd},  a pole  corresponding to  $P^{\Lambda}_{\psi
s}(4338)$ and an additional pole $P^{\Lambda}_{\psi s}(4254)$ near the
$\Lambda_c \bar{D}_s$ threshold were found important to reproduce the
experimental data.  However, in Ref.~\cite{Burns:2022uha}, the sharp
$P^{\Lambda}_{\psi s}(4338)$ enhancement is due to the triangle singularity in
another diagram featuring a $1/2^-$ baryon consistent with $\Sigma_c(2800)$.  In
Ref.~\cite{Meng:2022wgl}, the author studied the double thresholds distort the
line shapes of the $P^{\Lambda}_{\psi s}(4338)$ resonance in depth, and pointed
out that it is misleading to depict the line shapes with Breit-Wigner distribution.

Within a quasipotential Bethe-Salpeter equation (qBSE) approach,  possible
molecular states from the $\Xi_{c}^{(',*)}\bar{D}^{(*)}$ interactions were
studied in our previous work~\cite{Zhu:2021lhd}. The mass of the molecular state
$\Xi_c \bar{D}^{*} (3/2^-)$ is very close to the mass of $P^\Lambda_{\psi
s}(4459)$, but its widths is narrower than the experimental results.  The
results do not exclude the possibility of two-pole structure from $\Xi_c
\bar{D}^{*}$ states  with $1/2^-$ and $ 3/2^-$ and the role of either state in reproducing the
experimental  invariant mass spectrum.  It is worth mentioning that our previous
work also predicts other partners of $P^\Lambda_{\psi s}(4459)$ from the
$\Xi^{(',*)}_c \bar{D}^{(*)}$ interaction, including a molecular state from the
$\Xi_c \bar{D}$ interaction with $1/2^-$~\cite{Zhu:2021lhd}, which has a mass
close to the $P^{\Lambda}_{\psi s}(4338)$ observed in the recent LHCb
experiment~\cite{Collaboration:2022boa}.

In the previous work, only the $\Xi^{(',*)}_c \bar{D}^{(*)}$ channels which can
produce bound states were considered, and the coupling effects of unbound
channels were not included, especially the $\Lambda J/\psi$ channel where the
strange hidden-charm pentaquarks $P^{\Lambda}_{\psi s}(4459)$ and 
$P^{\Lambda}_{\psi s}(4338)$ were observed.  In the current work, a
coupled-channel calculation will be performed to estimate roles of the molecular
states in the $J/\psi\Lambda$  invariant mass spectrum and their relations to
the observed structures. To obtain more reasonable mass and width,  the full
coupled-channels effects for the $P^{\Lambda}_{\psi s}$ states are completely
considered, including channels $\Xi_c^{(',*)}\bar{D}^{(*)}$, $\Lambda_c\bar{D}_s^{(*)}$,
and  $\Lambda J/\psi$.  By comparison with experimental data, the origins of
the $P^{\Lambda}_{\psi s}(4459)$ and $P^{\Lambda}_{\psi s}(4338)$ will be
discussed.

After the introduction, Lagrangians used to construct the potential of
couple-channel  interaction,  the qBSE approach, and formula of  the invariant
mass spectrum will be presented in Sec.~\ref{Sec:
Formalism}.  In Sec.~\ref{sec3}, single-channel calculation
results, coupled-channel calculation results, and the estimations of the
$J/\psi\Lambda$  invariant mass spectrums will be given and discussed,
respectively.  The Sec.~\ref{sum} is a summary of the whole work and some
suggestions for experiment.

\section{Formalism}\label{Sec: Formalism}

We will search for the poles in the complex energy plane
within qBSE approach, and compare the results with  the $J/\psi\Lambda $
invariant mass spectrum. Hence, the potential kernel should be constructed first
with effective Lagrangians to calculate the scattering amplitudes, which are  used
to found the poles and estimate the invariant mass spectra.

\subsection{Relevant Lagrangians and Potentials}

In the current work, we consider all hidden-charm channels relevant to  the $P_{\psi
s}^\Lambda$, which explicitly  include $\Xi_c^*\bar{D}^*$, $\Xi'_c\bar{D}^*$,
$\Xi^*_c\bar{D}$, $\Xi_c\bar{D}^*$, $\Xi'_c\bar{D}$, $\Lambda_c\bar{D}_s^{*}$,
$\Xi_c\bar{D}$, $\Lambda_c\bar{D}_s$, and $\Lambda J/\psi$.  For the former eight
channels, we need the Lagrangians under the heavy quark limit and chiral
symmetry~\cite{Cheng:1992xi,Yan:1992gz,Wise:1992hn,Casalbuoni:1996pg,Zhu:2021lhd},
\begin{align}
  \mathcal{L}_{\mathcal{\tilde{P}}^*\mathcal{\tilde{P}}\mathbb{P}} &=
 i\frac{2g\sqrt{m_{\mathcal{\tilde{P}}} m_{\mathcal{\tilde{P}}^*}}}{f_\pi}
  (-\mathcal{\tilde{P}}^{*\dag}_{a\lambda}\mathcal{\tilde{P}}_b
  +\mathcal{\tilde{P}}^\dag_{a}\mathcal{\tilde{P}}^*_{b\lambda})
  \partial^\lambda\mathbb{P}_{ab},\nonumber\\
    \mathcal{L}_{\mathcal{\tilde{P}}^*\mathcal{\tilde{P}}^*\mathbb{P}} &=
-\frac{g}{f_\pi} \epsilon_{\alpha\mu\nu\lambda}\mathcal{\tilde{P}}^{*\mu\dag}_a
\overleftrightarrow{\partial}^\alpha \mathcal{\tilde{P}}^{*\lambda}_{b}\partial^\nu\mathbb{P}_{ba},\nonumber\\
    \mathcal{L}_{\mathcal{\tilde{P}}^*\mathcal{\tilde{P}}\mathbb{V}} &=
\sqrt{2}\lambda g_V\varepsilon_{\lambda\alpha\beta\mu}
  (-\mathcal{\tilde{P}}^{*\mu\dag}_a\overleftrightarrow{\partial}^\lambda
  \mathcal{\tilde{P}}_b  +\mathcal{\tilde{P}}^\dag_a\overleftrightarrow{\partial}^\lambda
 \mathcal{\tilde{P}}_b^{*\mu})(\partial^{\alpha}\mathbb{V}^\beta)_{ab},\nonumber\\
	\mathcal{L}_{\mathcal{\tilde{P}}\mathcal{\tilde{P}}\mathbb{V}} &= i\frac{\beta	g_V}{\sqrt{2}}\mathcal{\tilde{P}}_a^\dag
	\overleftrightarrow{\partial}_\mu \mathcal{\tilde{P}}_b\mathbb{V}^\mu_{ab}, \nonumber\\
  \mathcal{L}_{\mathcal{\tilde{P}}^*\mathcal{\tilde{P}}^*\mathbb{V}} &= - i\frac{\beta
  g_V}{\sqrt{2}}\mathcal{\tilde{P}}_a^{*\dag}\overleftrightarrow{\partial}_\mu
  \mathcal{\tilde{P}}^*_b\mathbb{V}^\mu_{ab}\nonumber\\
  &-i2\sqrt{2}\lambda  g_Vm_{\mathcal{\tilde{P}}^*}\mathcal{\tilde{P}}^{*\mu\dag}_a\mathcal{\tilde{P}}^{*\nu}_b(\partial_\mu\mathbb{V}_\nu-\partial_\nu\mathbb{V}_\mu)_{ab}
,\nonumber\\
  \mathcal{L}_{\mathcal{\tilde{P}}\mathcal{\tilde{P}}\sigma} &=
  -2g_s m_{\mathcal{\tilde{P}}}\mathcal{\tilde{P}}_a^\dag \mathcal{\tilde{P}}_a\sigma, \nonumber\\
  \mathcal{L}_{\mathcal{\tilde{P}}^*\mathcal{\tilde{P}}^*\sigma} &=
  2g_s m_{\mathcal{\tilde{P}}^*}\mathcal{\tilde{P}}_a^{*\dag}
  \mathcal{\tilde{P}}^*_a\sigma,\label{LD}
\end{align}
where  the $\mathcal{\tilde{P}}=(\bar{D}^0, D^-, D^-_s)$, and  the $\mathbb P$ and $\mathbb V$ are the pseudoscalar and vector matrices as
\begin{align}
    {\mathbb P}&=\left(\begin{array}{ccc}
        \frac{\sqrt{3}\pi^0+\eta}{\sqrt{6}}&\pi^+&K^+\\
        \pi^-&\frac{-\sqrt{3}\pi^0+\eta}{\sqrt{6}}&K^0\\
        K^-&\bar{K}^0&-\frac{2\eta}{\sqrt{6}}
\end{array}\right),
\mathbb{V}&=\left(\begin{array}{ccc}
\frac{\rho^{0}+\omega}{\sqrt{2}}&\rho^{+}&K^{*+}\\
\rho^{-}&\frac{-\rho^{0}+\omega}{\sqrt{2}}&K^{*0}\\
K^{*-}&\bar{K}^{*0}&\phi
\end{array}\right).\label{MPV}
\end{align}
where the indices $a, b=1, 2, 3$ are used to label the particle elements in the matrices ${\mathbb P}, {\mathbb V}$ and vector $\mathcal{P}$.

Then, the Lagrangians for the couplings between charmed baryon and light mesons can also be given as,

\begin{align}
{\cal L}_{BB\mathbb{P}}&=-i\frac{3g_1}{4f_\pi\sqrt{m_{\bar{B}}m_{B}}}~\epsilon^{\mu\nu\lambda\kappa}\partial_\nu \mathbb{P}~
\sum_{i,j=0,1}\bar{B}_{i\mu} \overleftrightarrow{\partial}_\kappa B_{j\lambda},\nonumber\\
{\cal L}_{BB\mathbb{V}}&=-\frac{\beta_S g_V}{2\sqrt{2m_{\bar{B}}m_{B}}}\mathbb{V}^\nu
 \sum_{i,j=0,1}\bar{B}_i^\mu \overleftrightarrow{\partial}_\nu B_{j\mu}\nonumber\\
&-\frac{\lambda_S
g_V}{\sqrt{2}}(\partial_\mu \mathbb{V}_\nu-\partial_\nu \mathbb{V}_\mu) \sum_{i,j=0,1}\bar{B}_i^\mu B_j^\nu,\nonumber\\
{\cal L}_{BB\sigma}&=\ell_S\sigma\sum_{i,j=0,1}\bar{B}_i^\mu B_{j\mu},\nonumber\\
    {\cal L}_{B_{\bar{3}}B_{\bar{3}}\mathbb{V}}&=-\frac{g_V\beta_B}{2\sqrt{2m_{\bar{B}_{\bar{3}}}m_{B_{\bar{3}}}} }\mathbb{V}^\mu\bar{B}_{\bar{3}}\overleftrightarrow{\partial}_\mu B_{\bar{3}},\nonumber\\
{\cal L}_{B_{\bar{3}}B_{\bar{3}}\sigma}&=i\ell_B \sigma \bar{B}_{\bar{3}}B_{\bar{3}},\nonumber\\
{\cal L}_{BB_{\bar{3}}\mathbb{P}}
    &=-i\frac{g_4}{f_\pi} \sum_i\bar{B}_i^\mu \partial_\mu \mathbb{P} B_{\bar{3}}+{\rm H.c.},\nonumber\\
{\cal L}_{BB_{\bar{3}}\mathbb{V}}    &=\frac{g_\mathbb{V}\lambda_I}{\sqrt{2m_{\bar{B}}m_{B_{\bar{3}}}}} \epsilon^{\mu\nu\lambda\kappa} \partial_\lambda \mathbb{V}_\kappa\sum_i\bar{B}_{i\nu} \overleftrightarrow{\partial}_\mu
   B_{\bar{3}}+{\rm H.c.},
   \label{LB}
\end{align}
where the Dirac spinor operators with label $i, j=0,1$ are defined as,
\begin{align}
{B}_{0\mu}&\equiv -\sqrt{\frac{1}{3}}(\gamma_{\mu}+v_{\mu})\gamma^{5}B; \, \,  \,  B_{1\mu}\equiv B^{*ab}_{\mu},\nonumber\\
{\bar{B}}_{0\mu}&\equiv\sqrt{\frac{1}{3}}\bar{B}\gamma^{5}(\gamma_{\mu}+v_{\mu});\, \,  \, \bar{B}_{1\mu}\equiv \bar{B}^{*}_{\mu},
\end{align}
and  the charmed baryon matrices are defined as
\begin{align}
B_{\bar{3}}&=\left(\begin{array}{ccc}
0&\Lambda^+_c&\Xi_c^+\\
-\Lambda_c^+&0&\Xi_c^0\\
-\Xi^+_c&-\Xi_c^0&0
\end{array}\right),\quad
B=\left(\begin{array}{ccc}
\Sigma_c^{++}&\frac{1}{\sqrt{2}}\Sigma^+_c&\frac{1}{\sqrt{2}}\Xi'^+_c\\
\frac{1}{\sqrt{2}}\Sigma^+_c&\Sigma_c^0&\frac{1}{\sqrt{2}}\Xi'^0_c\\
\frac{1}{\sqrt{2}}\Xi'^+_c&\frac{1}{\sqrt{2}}\Xi'^0_c&\Omega^0_c
\end{array}\right), \nonumber\\
B^*&=\left(\begin{array}{ccc}
\Sigma_c^{*++}&\frac{1}{\sqrt{2}}\Sigma^{*+}_c&\frac{1}{\sqrt{2}}\Xi^{*+}_c\\
\frac{1}{\sqrt{2}}\Sigma^{*+}_c&\Sigma_c^{*0}&\frac{1}{\sqrt{2}}\Xi^{*0}_c\\
\frac{1}{\sqrt{2}}\Xi^{*+}_c&\frac{1}{\sqrt{2}}\Xi^{*0}_c&\Omega^{*0}_c
\end{array}\right).\label{MBB}
\end{align}

The masses of  particles involved in the calculation are chosen as suggested central values in the Review of  Particle Physics  (PDG)~\cite{Tanabashi:2018oca}. The mass of broad $\sigma$ meson is chosen as 500 MeV.
 The  coupling constants involved are listed in Table~\ref{coupling}.

\renewcommand\tabcolsep{0.16cm}
\renewcommand{\arraystretch}{1.2}
\begin{table}[h!]
\caption{The coupling constants adopted in the
calculation, which are cited from the literature~\cite{Chen:2019asm,Liu:2011xc,Isola:2003fh,Falk:1992cx}. The $\lambda$ and $\lambda_{S,I}$ are in the units of GeV$^{-1}$. Others are in the units of $1$.
\label{coupling}}
\begin{tabular}{cccccccccccccccccc}\toprule[1pt]
$\beta$&$g$&$g_V$&$\lambda$ &$g_{s}$&$\ell_S$\\
$0.9$&$0.59$&$5.9$&$0.56$ &$0.76$&$6.2$\\\hline
$\beta_S$&$g_1$&$\lambda_S$ &$\beta_B$&$\ell_B$ &$g_4$&$\lambda_I$\\
$-1.74$&$-0.94$&$-3.31$&$-\beta_S/2$&$-\ell_S/2$&$3g_1/{(2\sqrt{2})}$&$-\lambda_S/\sqrt{8}$ \\
\bottomrule[1pt]
\end{tabular}
\end{table}

However, for the couplings of  former eight channels and the lowest channel
$\Lambda J/\psi$, the heavy quark effective Lagrangians are not enough for
calculation. Here, we apply effective Lagrangians under the SU(4)
symmetry as~\cite{Mueller-Groeling:1990uxr,Molina:2008jw,Shen:2017ayv,Shen:2019evi},
\begin{eqnarray}\label{Lag:all}
\mathcal{L}_{BBP} &=& \frac{g_{BBP}}{m_P} \bar B \gamma^\mu \gamma^5 \partial_\mu P B, \nonumber \\
\mathcal{L}_{BBV} &=& - g_{BBV} \bar B \gamma^\mu V_\mu B, \nonumber \\
\mathcal{L}_{BB^*P} &=& \frac{g_{BB^*P}}{m_P} (\bar B^{*\mu} B + \bar B B^{*\mu})\partial_\mu P, \nonumber \\
\mathcal{L}_{BB^*V} &=& -i \frac{g_{BB^*V}}{m_V} (\bar B^{*\mu} \gamma^5 \gamma^\nu B - \bar B \gamma^5 \gamma^\nu B^{*\mu}) \nonumber \\
&& \times (\partial_\mu V_\nu - \partial_\nu V_\mu), \nonumber \\
\mathcal{L}_{PPV} &=& - g_{PPV}(P \partial_\mu P - \partial_\mu P P)V^\mu, \nonumber \\
\mathcal{L}_{VVP} &=& \frac{g_{VVP}}{m_V} \epsilon_{\mu \nu \alpha \beta} \partial^\mu V^\nu \partial^\alpha V^\beta P, \nonumber \\
\mathcal{L}_{VVV} &=& g_{VVV} < (\partial_\mu V_\nu -\partial_\nu V_\mu)V^\mu V^\nu >,
\end{eqnarray}
where the involved coupling constants are shown in  Table~\ref{SUK}.

\renewcommand\tabcolsep{0.19cm}
\renewcommand{\arraystretch}{1.35}
\begin{table}[h!]
\caption{The coupling constants determined with SU(4) symmetry, and $g_{B^*BV}= 16.03$, $g_{B^*BP}= 2.127$, $g_{BBV}=3.25 $, $g_{BBP}= 0.989$,  $g_{PPV}=3.02$, $g_{VVP}=-7.07$ and $g_{VVV}=2.298$~\cite{Shen:2019evi}. The values are  in the units of GeV.}\label{SUK}
\begin{tabular}{lrrlrr}\toprule[1.pt]
 Coupl. 				&Relation	& Values			 				&Coupl.			                                  &Relation 		&Values\\ \hline
$g_{\Xi_c^*\Lambda D^*}$       &$\frac{1}{2}g_{B^*BV}$&$8.015 $                           & $g_{\Xi_c^*\Lambda D}$  &$\frac{1}{2}g_{B^*BP}$  &$1.109 $ \\
$g_{\Xi'_c\Lambda D^*}$       &$-\sqrt{\frac{3}{2}}g_{BBV}$&$-3.98 $          & $g_{\Xi'_c\Lambda D}$       &$\frac{\sqrt{3}}{5\sqrt{2}}g_{BBP}$&$ 0.242$ \\
$g_{\Xi_c\Lambda D^*}$       &$-\frac{1}{\sqrt{2}}g_{BBV}$&$ -2.298$                           & $g_{\Xi_c\Lambda D}$  &$-\frac{\sqrt{3}}{5\sqrt{2}}g_{BBP}$  &$-0.242 $ \\
$g_{\Lambda_c\Lambda D_s^*}$       &$\sqrt{2}g_{BBV}$&$ 4.596$                           & $g_{\Lambda_c\Lambda D_s}$  &$\frac{3\sqrt{2}}{5}g_{BBP}$  &$0.839 $ \\
$g_{D^*J/\psi D^*}$&$g_{VVV}$ &$2.298$                                                                               &$g_{D_s^*J/\psi D_s^*}$&$g_{VVV}$ &$2.298$      \\
$g_{D^*D J/\psi }$&$g_{VVP}$ &$-7.07$                                                                               &$g_{D_s^* D_s J/\psi }$&$g_{VVP}$ &$-7.07$      \\
$g_{DDJ/\psi}$&$-\sqrt{2}g_{PP V}$ &$-4.27$                                                                               &$g_{ D_s D_s J/\psi}$&$-\sqrt{2}g_{PPV}$ &$-4.27$      \\
\bottomrule[1pt]\hline
\end{tabular}
\end{table}

With the vertices obtained from above Lagrangians, the potential of couple-channel  interaction can be constructed easily with the help of the standard Feynman rules.
Because nine channels are involved in the current work, it is tedious and fallible to give  explicit 81 potential elements and input them into code.
Instead, in this work, following the method in Refs.~\cite{He:2019rva}, we input vertices $\Gamma$ and  propagators $P$  into  the code directly.  The potential can be written as
\begin{equation}%
{\cal V}_{\mathbb{P},\sigma}=f_I\Gamma_1\Gamma_2 P_{\mathbb{P},\sigma}f(q^2),\ \
{\cal V}_{\mathbb{V}}=f_I\Gamma_{1\mu}\Gamma_{2\nu}  P^{\mu\nu}_{\mathbb{V}}f(q^2).\label{V}
\end{equation}
The propagators are defined as usual as
\begin{equation}%
P_{\mathbb{P},\sigma}= \frac{i}{q^2-m_{\mathbb{P},\sigma}^2},\ \
P^{\mu\nu}_\mathbb{V}=i\frac{-g^{\mu\nu}+q^\mu q^\nu/m^2_{\mathbb{V}}}{q^2-m_\mathbb{V}^2},
\end{equation}
where the form factor $f(q^2)$ is adopted to compensate the off-shell effect of exchanged meson as $f(q^2)=e^{-(m_e^2-q^2)^2/\Lambda_e^2}$
with $m_e$ and $q$  being the mass and  momentum of the exchanged  meson,  respectively. And $\Lambda_e$ is  a cutoff to suppress the on-shell effect of exchange meson. In the propagator of exchanged meson, we make a replacement $q^2\to -|q|^2$ to remove singularities as in Ref~\cite{Gross:2008ps}.
The $f_I$ is the flavor factor for certain meson exchange of certain interaction with  isospin $I=0$, and the explicit values are listed in Table~\ref{flavor factor}.
\renewcommand\tabcolsep{0.16cm}
\renewcommand{\arraystretch}{1.2}
\begin{table}[h!]
\caption{The flavor factors $f_I$ for certain meson exchanges of certain interaction with isospin $I$=0.  The vertex for three pseudoscalar mesons should be forbidden.  \label{flavor factor}}
\begin{tabular}{c|ccccc}\bottomrule[1pt]
& $\pi$&$\eta$&$\rho$ & $\omega$ & $\sigma$ \\\hline
$\bar{D}^{(*)}\Xi^{(',*)}_c \to \bar{D}^{(*)}\Xi^{(',*)}_c$  &$-{3}/{4}$ &$-{1}/{12}$ &$-{3}/{4}$&${1}/{4}$ & 1\\
$\bar{D}^{(*)}\Xi_c\to \bar{D}^{(*)}\Xi_c$               & & &$-{3}/{2}$&${1}/{2}$ & 2\\
$\bar{D}^{(*)}\Xi_c\to \bar{D}^{(*)}\Xi_c^{(',*)}$& $-{3\sqrt{2}}/{4}$&$-\sqrt{2}/{4}$&$-{3\sqrt{2}}/{4}$&$\sqrt{2}/{4}$ & \\\bottomrule[1pt]
& $K$&$ K^*$&$$ & $ $ & $$ \\\hline
$\bar{D}^{(*)}\Xi^{(',*)}_c\to \bar{D}_s^{(*)}\Lambda_c$  &$-1$ &$-1$ &$ $&$ $ & $$\\
$\bar{D}^{(*)}\Xi_c\to \bar{D}_s^{(*)}\Lambda_c$  &&$\sqrt{2}$ &$ $&$ $ & $$\\
\bottomrule[1pt]
&$D$ & $ D^* $ & $D_s $& $D_s^* $&$  $ \\\hline
$\bar{D}^{(*)}\Xi_c\to J/\psi\Lambda$  &$ -\sqrt{2} $&$-\sqrt{2}  $& &$$ & $$\\
$\bar{D}^{(*)}\Xi^{(',*)}_c\to J/\psi\Lambda$  &$ -\sqrt{2}$&$-\sqrt{2} $&$$ &$$ & $$\\
$\bar{D}_s^{(*)}\Lambda_c\to J/\psi\Lambda$  &$$ &$$ &$1 $&$1 $ & $$\\

\toprule[1pt]
\end{tabular}
\end{table}

\subsection{The qBSE approach}

The  Bethe-Salpeter equation is a  four-dimensional integral equation in the
Minkowski space, which is hard to solve directly. By using  spectator
quasipotential approximation and  partial-wave decomposition, the
four-dimensional equation can be reduced into a 1-dimensional equation with fixed
spin-parity $J^P$ as~\cite{He:2015mja,He:2014nya,He:2012zd},
\begin{align}
i{\cal M}^{J^P}_{\lambda'\lambda}({\rm p}',{\rm p})
&=i{\cal V}^{J^P}_{\lambda',\lambda}({\rm p}',{\rm
p})+\sum_{\lambda''}\int\frac{{\rm
p}''^2d{\rm p}''}{(2\pi)^3}\nonumber\\
&\cdot
i{\cal V}^{J^P}_{\lambda'\lambda''}({\rm p}',{\rm p}'')
G_0({\rm p}'')i{\cal M}^{J^P}_{\lambda''\lambda}({\rm p}'',{\rm
p}),\quad\quad \label{Eq: BS_PWA}
\end{align}
where the sum extends only over nonnegative helicity $\lambda''$.
The $G_0({\rm p}'')$ is reduced from the 4-dimensional  propagator under quasipotential approximation as  $G_0({\rm p}'')=\delta^+(p''^{~2}_h-m_h^{2})/(p''^{~2}_l-m_l^{2})$ with $p''_{h,l}$ and $m_{h,l}$ being the momenta and masses of heavy or light constituent particles.
The partial wave potential is defined as,
\begin{align}
{\cal V}_{\lambda'\lambda}^{J^P}({\rm p}',{\rm p})
&=2\pi\int d\cos\theta
~[d^{J}_{\lambda\lambda'}(\theta)
{\cal V}_{\lambda'\lambda}({\bm p}',{\bm p})\nonumber\\
&+\eta d^{J}_{-\lambda\lambda'}(\theta)
{\cal V}_{\lambda'-\lambda}({\bm p}',{\bm p})],
\end{align}
where $\eta=PP_1P_2(-1)^{J-J_1-J_2}$ with $P$ and $J$ being parity and spin for system, constitution 1 or 2. The initial and final relative momenta are chosen as ${\bm p}=(0,0,{\rm p})$  and ${\bm p}'=({\rm p}'\sin\theta,0,{\rm p}'\cos\theta)$. The $d^J_{\lambda\lambda'}(\theta)$ is the Wigner d-matrix and the integration of the amplitude is
\begin{eqnarray}
\sum_{\lambda'\lambda}\int d\Omega |{\cal M}_{\lambda'\lambda}({\bm p}',{\bm p})|^2=\sum_{J^P,\lambda'\geq0\lambda\geq0}|\hat{\cal M}^{J^P}_{\lambda'\lambda}({\rm p}',{\rm p})|^2. \label{Eq: amplitudes sum}
\end{eqnarray}
To solve the integral equation (\ref{Eq: BS_PWA}), we discretize the momenta ${\rm p}$,
${\rm p}'$, and ${\rm p}''$ by the Gauss quadrature with a weight $w({\rm
p}_i)$ and have the discretized qBSE of the form as~\cite{He:2015mja}
\begin{eqnarray}
{M}_{ik}
&=&{V}_{ik}+\sum_{j=0}^N{ V}_{ij}G_j{M}_{jk}.\label{Eq: matrix}
\end{eqnarray}
The propagator $G$ is of a form
\begin{eqnarray}
	G_{j>0}&=&\frac{w({\rm p}''_j){\rm p}''^2_j}{(2\pi)^3}G_0({\rm
	p}''_j), \nonumber\\
G_{j=0}&=&-\frac{i{\rm p}''_o}{32\pi^2 W}+\sum_j
\left[\frac{w({\rm p}_j)}{(2\pi)^3}\frac{ {\rm p}''^2_o}
{2W{({\rm p}''^2_j-{\rm p}''^2_o)}}\right],
\end{eqnarray}
where on-shell momentum ${\rm p}''_o=\lambda^{1/2}(W,M_1,M_2)/2W$ with $
\lambda(x,y,z)=[x^2-(y+z)^2][x^2-(y-z)^2]$ and $W$ being the total energy of the
system of two constituents.  We also adopt an exponential regularization  by introducing a form
factor into the propagator as $G_0({\rm p}'')\to G_0({\rm
p}'')\left[e^{-(p''^2_l-m_l^2)^2/\Lambda_r^4}\right]^2$ with $\Lambda_r$ being a
cutoff ~\cite{He:2015mja}.  For simply, we set the cutoff $\Lambda_e=\Lambda_r =\Lambda$ and adjust
the value of $\Lambda$ around $1$~GeV.  However, the key observation channel
$\Lambda J/\psi$ couples with other channels only depend on the charmed heavy
mesons like $D, D_s, D^*, D_s^*$.  The cutoff $\Lambda_{(D, D_s)}$ and
$\Lambda_{(D^*, D_s^*)}$ will be chosen at least larger than their masses.
Specific values will be mentioned later.

The poles of  nine-channel scattering amplitudes with different spin parities $J^P$ are
searched by variation of $z$ in the complex energy plane to satisfy $|1-V(z)G(z)|=0$.
For $N$ channels calculation, there are a total number of $2^N$ Riemann sheets related to
unitary. In the current work, the treatment in Ref.~\cite{Roca:2005nm} is adopted to search for
the poles. In such treatment, the default propagator $G(z)$ is adopted in energy
region below the threshold and the $G(z)+{i{\rm p}''_o}/{16\pi^2 z}$ for
region above threshold, which correspond to the first and second Riemann
sheets, respectively.

The amplitudes of $ \Lambda J/\psi$ scattering can be obtained at the same time.
In the current work, The invariant mass spectrum of $\Lambda J/\psi$ channel can
be given approximately with the scattering amplitude ${\cal M}^{J^P}_{J/\psi \Lambda}$ as
~\cite{Hyodo:2003jw,Aceti:2014uea},
\begin{align}
	\frac{d\Gamma}{dW}&=\sum_{J^P} C_{J^P}\sum_{\lambda'\geq0\lambda\geq0}|{\cal M}^{J^P}_{ J/\psi\Lambda, \lambda'\lambda}({\rm p}',{\rm p})|^2\nonumber\\
  &\cdot \lambda^\half(W,m_{J/\psi},m_\Lambda)\lambda'^\half
	(\tilde{W},W,m_{3})/W,  \label{Ims}
\end{align}
with $\tilde{W}$ being total energy of the decay process, that is, the mass of
$\Xi^-_b$ baryon or $B^-$ meson. For simplicity, we do not calculate the yield of initial
decay process from $\Xi^-_b$ baryon or $B^-$ meson , and a yield parameter $C_{J^P}$ for spin parity $J^P$ is
introduced to absorb all uncertainties and the information about the yield.

\section{Result and Discussion}\label{sec3}
\subsection{Bound states from single-channel calculation}

\begin{figure}[h!]
  \centering
  \includegraphics[scale=0.75,bb=80 40 440 320,clip]{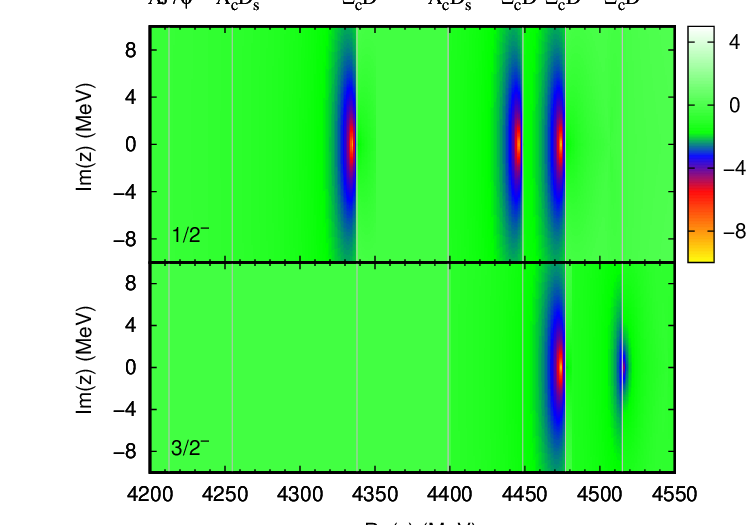}\\
  \caption{Poles with  spin parities $1/2^-$ (upper panel) and $3/2^-$ (lower panel) from the single-channel interactions.  The gray lines  correspond to the thresholds of the interactions. The color box is for the $\log|1-V(z)G(z)|$.}\label{1}
\end{figure}

The hadronic molecular state is a bound state from the interaction of two
hadrons. The single-channel calculation will provide the basic picture of
molecular states from the interaction considered. Here, a search for the poles
from  single-channel interactions will be first performed by the
$\log|1-V(z)G(z)|$ with the variation of real and imaginary parts of $z$. The
results with a cutoff $\Lambda=1.1$~GeV are presented in Fig.\ref{1}, where the
abscissa corresponds to the real part of the pole position, and the ordinate
corresponds to the imaginary part of the pole position. In this work, we only
consider the bound states which can be introduced in S wave. However, in the
calculation, higher-wave contributions for these states are included.

There are five poles found from the interactions considered in energy region from 4200 to 4550~MeV. Since only
single-channel interactions are considered, all poles are found on the real axis
of the complex energy plane. These poles correspond to the five molecular states
generated from the single-channel interaction of $\Xi_c^*\bar{D}$ with spin parity $J^P=3/2^-$,
$\Xi_c\bar{D}^*$ with $1/2^-$ and $3/2^-$,  $\Xi'_c\bar{D}$ with $1/2^-$, and
$\Xi_c\bar{D}$ with $1/2^-$. If the cutoff increases, all poles will leave
corresponding thresholds further.  Two states with  $1/2^-$ and
$3/2^-$ from the $\Xi_c\bar{D}^*$ interaction have almost the same mass with
single-channel calculation, about $4474$~MeV, which is a little larger than that
of $P^{\Lambda}_{\psi s}(4459)$.  They only can be distinguished after considering
coupling effects and the cutoff $\Lambda$ also should be adjusted a little
larger. The  $\Xi_c\bar{D}(1/2^-)$ interaction can produce an S-wave bound state
with a mass of about $4335$~MeV, which is very close to the mass of
$P^{\Lambda}_{\psi s}(4338)$. The  $\Xi'_c\bar{D}$ interaction produces a
molecular state with spin parity $1/2^-$. The $\Xi_c^*\bar{D}$ channel  produces
a molecular state with spin parity $3/2^-$, which has relatively weaker attraction
than the other four states.  Its mass is almost at the threshold, but the
binding energy will  increase with the increase of cutoff.  Since there is no
light meson exchange potential as the attractive mechanism, the rest three
channels $\Lambda_c\bar{D}_s^{*}$, $\Lambda_c\bar{D}_s$ and $\Lambda J/\psi$ cannot be bound
in the current model even if  the cutoff $\Lambda$ is adjusted to the largest
value in reasonable range.

\subsection{States near $\Xi_c\bar{D}^*$ threshold and $P^{\Lambda}_{\psi s}(4459)$}

\begin{figure}[h!]
  \centering
  \includegraphics[scale=0.9,bb=80 40 350 325,clip]{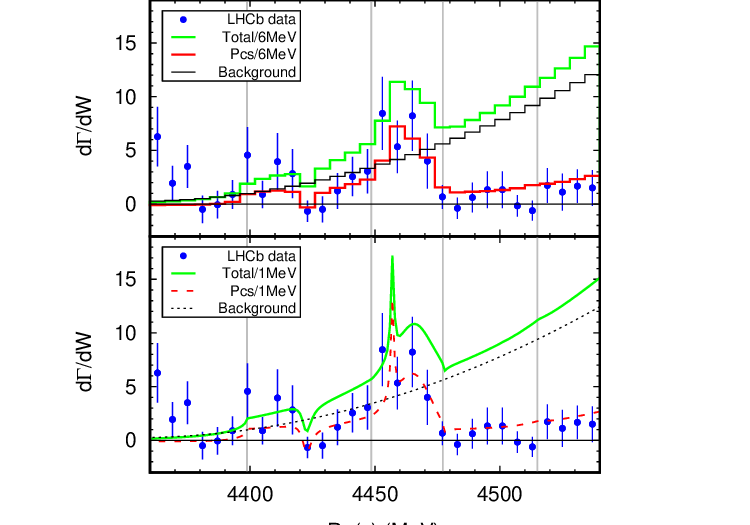}
  \caption{ The $J/\psi\Lambda $ invariant mass spectrums with bin of 6 MeV (upper panel) and of 1~MeV (lower panel). The green, red, and black curves are for the total, peak, and background contributions.  The blue points with error bars are the data removing the background contribution from the LHCb experiment~\cite{Aaij:2020gdg}. The gray lines  correspond to the thresholds of the interactions. }\label{mix16}
\end{figure}

For better understanding of the origin of the $P^{\Lambda}_{\psi s}(4459)$, the
coupled-channel effects will be included to introduce width to the bound states
and estimate the $J/\psi\Lambda$ invariant mass  spectrum in energy region of
$[4360-4540]$~MeV where the $P^{\Lambda}_{\psi s}(4459)$ was observed.  As shown
in the single-channel calculation, four molecular states are produced in this energy region, including $\Xi'_c\bar{D}(1/2^-)$, $\Xi_c\bar{D}^*(1/2^-, 3/2^-)$, and
$\Xi_c^*\bar{D}(3/2^-)$. The yield parameters $C_{1/2^-}$ and  $C_{3/2^-}$ are set completely free. To compare with the experimental invariant mass
spectrum, the corresponding cutoff $\Lambda$ will be adjusted to $1.14$~GeV,
which is a little larger than the one in single-channel calculation, to move the
poles to the experimental peak. The cutoff $\Lambda_{(D, D_s)}$ and
$\Lambda_{(D^*, D_s^*)}$  will be set as $2.25$ and $2.4$~GeV, which do not
involve in single-channel calculation.  With the increase of the cutoff
$\Lambda$,  the binding of states becomes deeper accordingly. The estimated
invariant mass spectrum is presented in Fig.~\ref{mix16} and compared with the
experiment~\cite{Aaij:2020gdg}.

In the upper panel of Fig.~\ref{mix16}, the theoretical results with a bin of 6~MeV
are presented, and compared with the LHCb experiment with the same
bin~\cite{Aaij:2020gdg}. Besides an obvious peak from the molecular states, the
direct results from the current coupled-channel calculation provide a background
contributions in the $J/\psi\Lambda$ invariant mass spectrum (see green curves
in Fig.~\ref{mix16}).  It is simply removed as a unary quadratic function
$BK(W)=6646.84-3053.5W+350.7W^2$ by comparing the theoretical and experimental results in
the energy region of $[4360-4390]\cup[4480-4540]$~MeV (see black curves in
Fig.~\ref{mix16}).  After removing such background, the results from the
molecular states are extracted, and in good line with the experimental data.  To
find the best comparison with experiment with bin of 6~MeV, the contributions of
two $\Xi_c\bar{D}^*$ states with spin parity $1/2^-$ and $3/2^-$ exhibit as
one peak.  If we adjust bin to a smaller value, 1~MeV, in the lower panel of
Fig.~\ref{mix16},  the contributions from the two states are distinguishable.
It shows that the peak of $P^{\Lambda}_{\psi s}(4459)$ may be  composed of a
higher narrow $\Xi_c\bar{D}^*$ state with $J^P=3/2^-$ and a lower wider $\Xi_c\bar{D}^*$
state with $J^P=1/2^-$.

\begin{figure}[h!]
  \centering
  \includegraphics[scale=0.82,bb=80 40 410 285,clip]{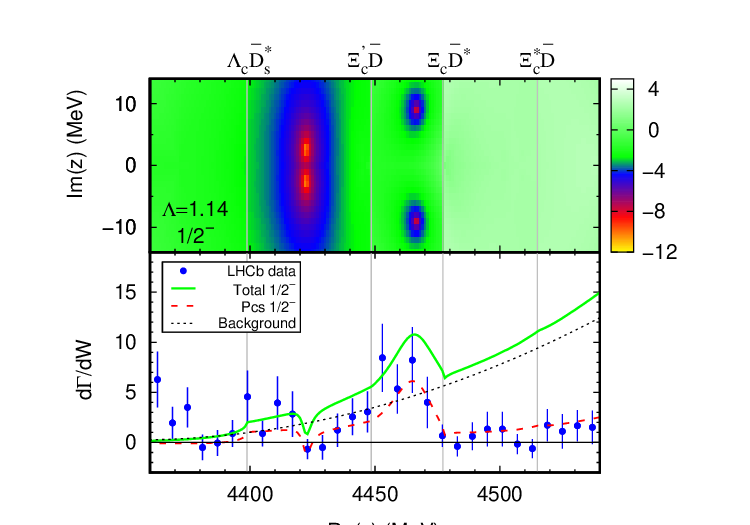}\\
  \caption{The poles (upper panel) and the  $ J/\psi\Lambda$ invariant mass spectrum (lower panel) for spin parity $1/2^-$. The green (full), red (dashed), and black  (dotted) curves are for the total, peak, and background contributions.  The blue points with error bars are the data removing the background contribution from the LHCb experiment cited from Ref.~\cite{Aaij:2020gdg}. The gray lines  correspond to the thresholds of the interactions. }\label{Fig:4459.1}
\end{figure}

To provide more explicit of the contributions from these two $\Xi_c\bar{D}^*$
states, the poles and the invariant mass spectrum for different spin parities
will be presented. The results for spin party  $1/2^-$ is illustrated in
Fig.~\ref{Fig:4459.1}.
In the single-channel calculation, where two poles are
found on the real axis near the $\Xi'_c\bar{D}$ and $\Xi_c\bar{D}^*$thresholds,
respectively.  After including the coupled-channel effect, the poles leave the
real axis and become two conjugate  poles in the complex energy planes, which
indicates that the bound states acquire width. The $\Xi_c\bar{D}^*$ state with
spin parity $1/2^-$ has a mass of  $4465$~MeV and a width of $18$~MeV.  The
structure $P^{\Lambda}_{\psi s}(4459)$ peak can even be well describe by the
peak of state $\Xi_c\bar{D}^*(1/2^-)$, except for the data points at about
$4453$~MeV. The $\Xi'_c\bar{D}(1/2^-)$ is a relative narrow state at about
$4423$~MeV with a width of about $6$~MeV.  In the experimental spectrum, there
is a small dip around $4423$~MeV, which can be in good conformity with the
molecular state $\Xi'_c \bar{D}(1/2^-)$. Unfortunately,  precision of the
experiment are not high enough to identify its existence.  The calculation also
suggests to observe such state in the $\Lambda_c \bar{D}_s$ invariant mass
spectrum.

\begin{figure}[h!]
  \includegraphics[scale=0.82,bb=80 40 410 285,clip]{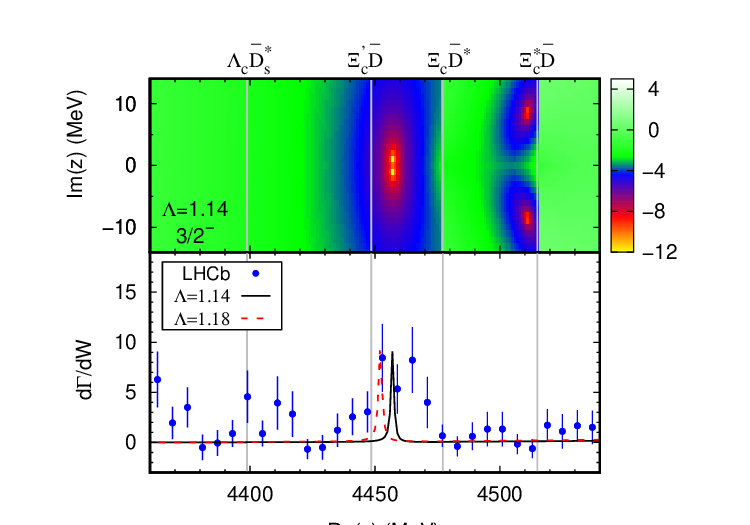}\\
  \caption{ The poles (upper panel)  and $J/\psi\Lambda $ invariant mass spectrum (lower panel) for spin parity $1/2^-$.  The black (full) and red (dashed) lines are for the peak from the coupled-channel interactions with cutoff 1.14 and 1.18, respectively. The blue points with error bars are the data removing the background contribution from LHCb experiment~\cite{Aaij:2020gdg}. The gray lines  correspond to the thresholds of the interactions.}\label{Fig:4459.2}
\end{figure}

In Fig.~\ref{Fig:4459.2}, the results for spin parity $3/2^-$ are presented. The
$\Xi_c\bar{D}^*$ state with $3/2^-$, as well as that with $1/2^-$, is produced
near the threshold with a mass of $4457$~MeV. The width of the state
$\Xi_c\bar{D}^*(3/2^-)$ equals to about $1.6$~MeV, being  much smaller than that
of the state with $1/2^-$.  The peak of the state $\Xi_c\bar{D}^*(3/2^-)$ is so
narrow that it can only raise the events near the central value and not
affect much of the general shape of the invariant mass spectrum.  The peak
with cutoff $\Lambda=1.14$ falls in the middle of two highest data points. If a little larger 
cutoff $\Lambda=1.18$ is adopted, the peak will move
to the lower highest data point.  The state $\Xi^*_c\bar{D}(3/2^-)$ has a mass of
about $4512$~MeV and a width of about $17$~MeV.  The low yield of the events with
$J^P=3/2^-$ indicates that high precision data are required to observe the
$\Xi^*_c\bar {D} (3/2^-)$  in $J/\psi\Lambda $ invariant mass spectrum. The
decay channels $\Xi_c\bar{D}^*$ and $\Lambda_c \bar{D}_s^*$ couple strongly with
the state, which will be good choices to search for this state.

\subsection{State near $\Xi_c\bar{D}$ threshold and $P^{\Lambda}_{\psi
s}(4338)$}

The $P^{\Lambda}_{\psi s}(4338)$ was observed near the $\Xi_c\bar{D}$ threshold
in the process $B^-$$\to$$J/\psi \Lambda \bar{p}$. The single-channel results
suggest a bound state from the  $\Xi_c\bar{D}$ interaction, which is close to
the $P^{\Lambda}_{\psi s}(4338)$. To confirm their relation, the invariant mass
spectrum is estimated and compared with the experiment.  To match the
experimental data points, we need to introduce a parametrized background
contribution of a form $M^{BK,J^P}_{\Lambda J/\psi}=a e^{ib\pi}(M_{\Lambda
J/\psi}-M_{min})^c(M_{max}-M_{\Lambda J/\psi})^d$ with the $M_{max} $ and
$M_{min}$ being the upper and lower limits of phase space, which interferes with
the amplitude from the $\Xi_c\bar{D}$  interaction. Considering that the
$P^{\Lambda}_{\psi s}(4338)$ is very close to the threshold, the cutoff
$\Lambda$ is adjusted to $1.04$~GeV,  a little smaller than $1.1$~GeV, and the
cutoff $\Lambda_{(D, D_s)}$ and  $\Lambda_{(D^*, D_s^*)}$ will be set as
$2.15$ and $2.3$~GeV. The parameters for the background will be set as $(a,
b, c, d)=(0.054, 0.19, 0.085, 0.28)$. The yield parameter $C_{1/2^-}$  is set completely free.  The results are shown in the following
Fig.~\ref{Fig:4338} and compared with recent LHCb experiment.

\begin{figure}[h!]
  \centering
  \includegraphics[scale=0.82,bb=75 10 380 285,clip]{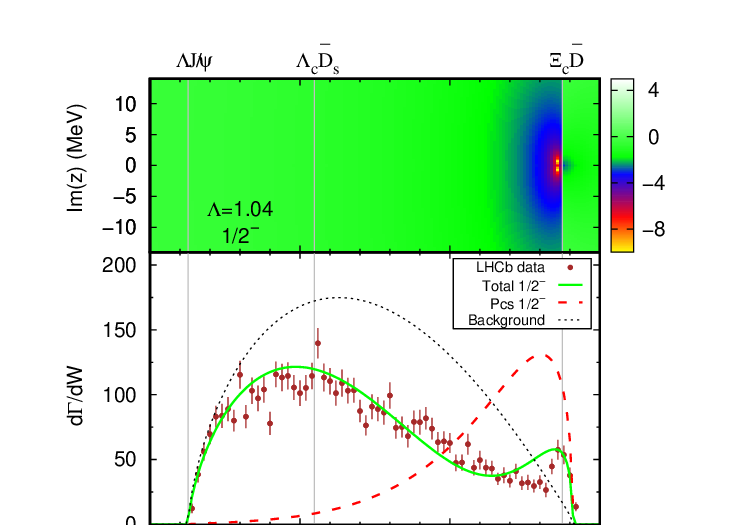}\\
  \caption{ The poles (upper panel)  and $ J/\psi\Lambda$ invariant mass spectrum (lower panel) for spin parity $1/2^-$.  The green (full), red (dashed), and black  (dotted) curves are for the total, peak, and background contributions.   The brown points with error bars are the data removing the background contribution from the LHCb experiment~\cite{Collaboration:2022boa}. The gray lines  correspond to the thresholds of the interactions. }\label{Fig:4338}
\end{figure}

The diagram of poles shows that a very narrow molecular state from the
$\Xi_c\bar{D}$ interaction with spin parity $1/2^-$, which is very close to the
corresponding threshold.  The mass and width of the state are about $4336.5$~MeV
and $0.8$~MeV, respectively.  The strong coupling between the $\Lambda_c
\bar{D}_s$ and the $\Xi_c\bar{D}$ channels provides the dominant contribution of
its width.  In the $J/\psi\Lambda $ invariant mass spectrum,  the angle of
interference between them has reached about $34.2^\circ$. Since the pole is very close to the real axis, a very narrow peak can be produced from the square of the amplitudes $|{\cal M}^{J^P}_{ J/\psi\Lambda, \lambda'\lambda}({\rm p}',{\rm p})|^2$ in Eq.~(\ref{Ims}). However, the pole
produced from the $\Xi_c\bar{D}$ interaction is close to the threshold of the
phase space. The original narrow peak is suppressed into  a relatively wide peak by the factor $\lambda'(\tilde{W},W,m_{3})$. Such a
result implies that the width of $P^{\Lambda}_{\psi s}(4338)$ determined
experimentally need more analysis.  An obvious enhancement can be produced from
the contribution of molecular state $\Xi_c\bar{D}(1/2^-)$.  Obviously,  the structure $P^{\Lambda}_{\psi s}(4338)$ is 
 in line with the molecular state of  $\Xi_c\bar{D}(1/2^-)$. The peak
from the molecular state seems obvious wider than the experimental structure. 
 In Ref.~\cite{Nakamura:2022jpd} , more contributions, such as threshold cusp effects, are included.  The  line shape in the
no-pole model suggests the $\Xi_c \bar{D}$ threshold cusp plays important
role in the $P^\Lambda_{\psi s}(4338)$ peak structure.  
The inclusion of those contributions may be helpful to better understand the sharp peak structure. it seems that more elaborate structures like $\Xi_c^+ D^-$ and $\Xi_c^0
\bar{D}^0$ channels under the uncoupled isospin representation is also helpful to obtain n sharp peak as  discussed in Ref.~\cite{Meng:2022wgl}. Besides, if the  strong coupling between  $\Lambda_c \bar{D}_s$ and $\Xi_c\bar{D}$ channels is weakened, the broad bump can also become a narrow peak. However, in the current work, there is only one adjustable  parameter  $\Lambda$,  it is impossible to weaken the coupling of certain two channels but keep strengths of other interactions unchanged.

\section{Summary}\label{sum}

In this work, we perform a coupled-channel calculation to study the molecular
states produced from interactions $\Xi_c^*\bar{D}^*$,
$\Xi'_c\bar{D}^*$,$\Xi^*_c\bar{D}$, $\Xi_c\bar{D}^*$, $\Xi'_c\bar{D}$,
$\Lambda_c\bar{D}_s^{*}$, $\Xi_c\bar{D}$, $\Lambda_c\bar{D}_s$, and $\Lambda J/\psi$. The
poles of the  molecular states are searched in complex energy plane in the
qBSE approach. With the help of effective
Lagrangian, the potential kernel can be constructed by meson exchanges. With the
scattering amplitudes obtained, the invariant mass spectra are estimated and
compared with the experiment. Based on the current results, the understanding of
the experimentally observed $P^{\Lambda}_{\psi s}(4459)$ and $P^{\Lambda}_{\psi
s}(4438)$, as well as their partners, can be drawn as follows. 

\begin{itemize}

  \item The $\Xi_c \bar{D}^{*}$ state with spin parity $1/2^-$ can reproduce the
  general line shape of the structure $P^{\Lambda}_{\psi s}(4459)$ in the
  $J/\psi\Lambda$ invariant mass spectrum.  However, the contribution from the
  $\Xi_c \bar{D}^{*}$ state with $3/2^-$, which is lower than the $1/2^-$ state, cannot be excluded due to its very
  small width. The high-precision data is required to understand its role in
  this structure.

  \item Based on the $J/\psi\Lambda$ invariant mass spectrum, the existence of
  $P^{\Lambda}_{\psi s}(4459)$ suggests a dip near the $\Xi'_c \bar{D}$
  threshold in the LHCb experimental data. Such dip is consistent with a $\Xi'_c \bar{D}$ molecular state with $1/2^-$ with a  mass of $4423$~MeV and a width of  $6$~MeV. The   $\Lambda_cD_s$ channel is a good place to search for  this state.

  \item  The state $\Xi^*_c \bar{D}(3/2^-)$  can be produced from the coupled
  channel calculation with a mass of $4512$~MeV and a  width of $17$~MeV.
  However, both  experimental and theoretical results suggest that it couples
  very weakly to the $\Lambda J/\psi$  channel. It is suggested to search for
  such state in the channels $\Xi_c\bar{D}^*$ and $\Lambda_cD_s^{*}$.

  \item A molecular state with a very small width can be produced from the
  $\Xi_c \bar{D}$ interaction with spin parity $(1/2^-)$, which is very close to the $P^{\Lambda}_{\psi s}(4338)$.  Obviously, the structure $P^{\Lambda}_{\psi s}(4338)$ is 
 in line with the molecular state of  $\Xi_c\bar{D}(1/2^-)$.  The phase
  space suppress it into a wider peak near the threshold. Large
  interference with background should be introduced to reproduce a narrow peak. The $\Xi_c\bar{D}$ threshold cusp effect may be also important to cause the peak~\cite{Nakamura:2022jpd}.
  In addition, the molecular state $\Xi_c \bar{D}(1/2^-)$ also can be search
  for in the $\Lambda_cD_s$ channel.

  \item  Within current model, there is no bound state produced from the
  interactions   $\Lambda_c\bar{D}^*_s$, $\Lambda_c\bar{D}_s$, and $\Lambda
  J/\psi$.
  
 \end{itemize}

The current calculation is performed with only the contribution of  molecular states from coupled-channel interactions and  a parametrized background. Other contributions, such as the cusp effects,  are not included.  The initial decay processes are also parametrized as free parameters.  To better understand the roles of the molecular states  in such decays and their relations to the observed structures, inclusion of more contributions will be helpful.

\section{Acknowledgement}
This project is supported by the Postgraduate Research
\& Practice Innovation Program of Jiangsu Province (Grants No. KYCX21\_1323)) and the National Natural Science
Foundation of China (Grants No. 11675228).

\end{document}